\documentclass{aastex}
\usepackage{spr-astr-addons}

\begin{document}

\title{Constraints on Spin of a Supermassive Black Hole in Quasars with Big Blue Bump \\
{\small Accepted for publication in Astrophysics and Space Science}}

\author{M. Yu. Piotrovich} \and \author{Yu. N. Gnedin} \and \author{T. M. Natsvlishvili} \and \author{S. D. Buliga}
\affil{Central Astronomical Observatory at Pulkovo, Saint-Petersburg, Russia.}
\email{mpiotrovich@mail.ru}

\begin{abstract}
We determined the spin value of supermassive black hole (SMBH) in active galactic nuclei (AGN) with investigated ultraviolet-to-optical spectral energy distribution, presented in the sample of Shang et al. (2005). The estimates of the spin values have been produced at the base of the standard geometrically thin accretion disk model and with using the results of the polarimetric observations. The polarimetric observations are very important for determining the inclination angle of AGN disk. We presented the results of our determinations of the radiation efficiency of the accretion flow and values of the spins of SMBHs, that derives the coefficient of radiation efficiency. The majority of SMBHs of AGNs from Shang et al. (2005) sample are to be the Kerr black holes with the high spin value.
\end{abstract}


\section{Introduction} 

It is accepted that AGNs are powered by accretion onto SMBHs. It is very important that the spins of SMBHs play a basic role in their growth and hole hints to their evolution. Also the spins of SMBHs can be an important factor for generation of relativistic jets. According to the standard accretion disk model \citep{shakura73} the dimensionless spin $a$ of the SMBH determines the radius of the innermost stable orbit in the accretion disk and the radiative efficiency of the accretion flow $\varepsilon (a) = L_{bol} / \dot{M} c^2$, where $\dot{M}$ is the accretion rate and $L_{bol}$ is the bolometric luminosity of the accretion disk.

One of the most important method for measuring the spin is the analysis of the relativistically broadened X-ray spectral lines from the inner accretion disk that provides an effective tool for measuring the spin of SMBH in AGN \citep{brenneman06,reynolds14,brenneman13,brenneman13b}. At the base of X-ray spectral line profile it is possible to derive the value of the radius of the innermost stable circular orbit (ISCO) in the accretion disk. Thus value is strongly dependent on the black hole spin \citep{reynolds14,brenneman13}. However, the profiles of X-ray spectral lines are complex and their interpretation requires the detail modeling.

Other useful method for measuring the spin is based on estimates of the kinetic power of the relativistic jets which is depending strongly on the spin value \citep{daly11}. Estimates of the kinetic power of relativistic jets are obtained using some estimate of the magnetic field strength near the event horizon radius of a SMBH if the Blandford-Znajek mechanism is assumed \citep{blandford77} or near the radius of the ISCO in the accretion disk \citep{blandford82,garofalo10}. In this situation the determination of the magnetic field is the central problem.

Therefore it is useful to test the various methods for determining the spin value of SMBH. Traditionally the spin value is estimated in terms of the dimensionless parameter $a = c J / G M_{BH}^2$, where $J$ is the angular momentum and $M_{BH}$ is the mass of the black hole. The spin can take on values $0 \leq a \leq 0.998$ \citep{thorne74}, but it can have a negative value, i.e. $a < 0$. The negative value corresponds to the situation when the direction of the central black hole rotation is retrograde concerning the rotation of the accretion disk. For example, \citet{garofalo10} have suggested that the jets in the most energetic radio galaxies may be powered by accretion onto rapidly rotating retrograde BHs. This may be of sone importance for understanding the properties of powerful radio loud AGNs. The accretion disk around the retrograde black hole is potent configuration for generating powerful jets.

The effective method of the black hole spin determining is connected with the determination of the radiative efficiency of AGN $\varepsilon (a)$ that derives the conversion process of gravitational energy into radiation and depends strongly on the spin of SMBH \citep{novikov73,davis06,krolik07,krolik07b,schnittman16}:

\begin{equation}
 \varepsilon (a) = \frac{L_{bol}}{\dot{M}c^2},
 \label{eq01}
\end{equation}

\noindent where $L_{bol}$ is the bolometric luminosity and $\dot{M}$ is the accretion rate.

\citet{collin02,collin06} and \citet{davis11} estimated the radiative efficiency of a number of AGNs. Unfortunately, they suggested the definite value of the inclination angle for the accretion disk. For example, \citet{davis11} suggested that the value of cosine of inclination angle $i$ is $\mu = \cos{i} = 0.8$.

\section{Basic equations}

In the standard geometrically thin accretion disk \citep{shakura73} the rest frame continuum emission of this disk follows a power-law form $L_\nu \sim (M_{BH}\dot{M})^{2/3} \nu^{1/3}$ \citep{davis11}. Using this relation \citet{du14} have obtained the following expression:

\begin{equation}
 \mu^{3/2} l_E = 0.201 \left(\frac{L_{5100}}{10^{44} erg/s}\right)^{3/2} \frac{\varepsilon (a)}{M_8^2},
 \label{eq02}
\end{equation}

\noindent where $L_{5100} = \lambda L_\lambda (5100 \text{\AA})$, $l_E = L_{bol} / L_{Edd}$ is the Eddington ratio and $M_8 = M_{BH} / 10^8 M_{\odot}$. The similar result was obtained by \citet{trakhtenbrot14}.

The equation (\ref{eq02}) determines the value of the radiation efficiency $\varepsilon (a)$ that is equal to the mass-to-energy conversion efficiency

\begin{equation}
 \varepsilon (a) = 1 - \left(1 - \frac{2 R_g}{3 R_{ISCO}}\right)^{1/2},
 \label{eq03}
\end{equation}

\noindent where $(1 - 2 R_g / 3 R_{ISCO})^{1/2}$ determines the specific energy of the accretion disk particle at the ISCO and $R_{ISCO}$ is the radius of this orbit \citep{bardeen72,novikov73}.

The relationship between the radiation efficiency $\varepsilon$ and the spin $a$ is presented at Fig.1.

\begin{figure}[t]
	\includegraphics[width=\columnwidth]{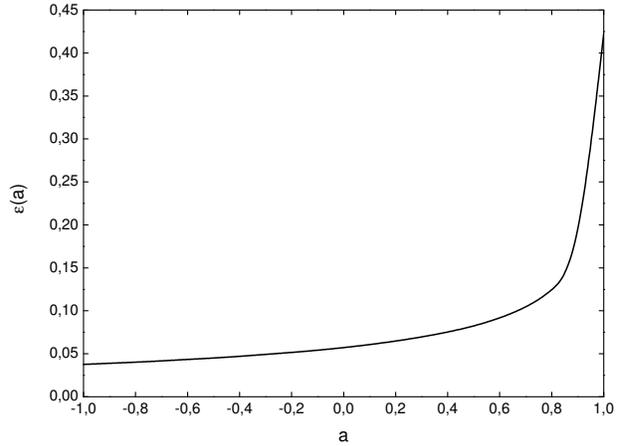}
  \caption{The relationship between the radiation efficiency $\varepsilon$ and the spin $a$}
  \label{fig01}
\end{figure}

\section{Sample and data for determining SMBH spins} 

\begin{table*}[t]
\small
\caption{Spin values of SMBHs for AGNs from \citet{shang05}, obtained with relation from \citet{du14}. $l_E$ is Eddington ratio, $\mu = \cos{i}$, $i$ is inclination angle, $P_l(\mu)$ is polarization degree, $a$ is value of spin. The value of $\cos{i}$ for (1) is obtained using Eqs.(\ref{eq05})-(\ref{eq07}), for (2) is obtained using polarimetric data from \citet{smith02}, for (3) is obtained from \citet{berriman90}, for (4) is obtained from \citet{afanasiev11}.}
\label{tab1}
\centering
\begin{tabular}{lccccc}
\tableline
{\bf Object}  & $l_E$ & $\mu$          & $P_l(\mu)$[\%] & $\varepsilon(a)$       & $a$ \\
\tableline
3C 273        & 0.41  & $\geq 0.865^1$& $\leq 0.44$    & $\leq 0.42$            & $\leq 1.0$ \\
3C 351        & 0.06  & $0.892^1$     & 0.33           & $\leq 0.32$            & $\leq 0.998$ \\
4C+34.37      & 0.30  & $0.900^1$     & 0.32           & $0.20 \pm 0.01$        & $0.95 \pm 0.01$ \\
IR 07546+3928 & 0.39  & $0.893^1$     & 0.34           & $0.137 \pm 0.012$      & $0.85 \pm 0.02$ \\
Mrk 506       & 0.05  & $0.825^1$     & 0.58           & $0.20^{+0.04}_{-0.03}$ & $0.95\pm 0.02$ \\
PG 1259+593   & 0.45  & $0.875^1$     & 0.40           & $0.31 \pm 0.02$        & $0.998 \pm 0.01$ \\
PG 2349-014   & 0.09  & $0.805^1$     & 0.66           & $< 0.42$               & $\leq 1.0$ \\
Mrk 290       & 0.04  & $0.747^2$     & 0.90           & $0.19 \pm 0.01$        & $0.95 \pm 0.01$ \\
Mrk 304       & 0.09  & $0.834^2$     & 0.55           & $0.35 \pm 0.05$        & $0.998 \pm 0.003$ \\
Mrk 509       & 0.16  & $0.762^2$     & 0.47-0.84      & $0.14 \pm 0.01$        & $0.84\pm 0.01$ \\
PG 0052+251   & 0.11  & $0.903^3$     & $0.30\pm 0.12$ & $0.31 \pm 0.02$        & $0.998\pm 0.002$ \\
PG 0947+396   & 0.22  & $0.887^3$     & 0.36           & $0.23 \pm 0.03$        & $0.96 \pm 0.02$ \\
PG 1100+772   & 0.06  & $0.793^3$     & 0.71           & $0.14^{+0.02}_{-0.01}$ & $0.88^{+0.02}_{-0.03}$ \\
PG 1322+659   & 0.32  & $0.776^3$     & 0.78           & $0.25 \pm 0.03$        & $0.98^{+0.015}_{-0.02}$ \\
PG 1351+640   & 0.43  & $0.805^3$     & 0.66           & $0.146^{+0.04}_{-0.06}$& $0.89^{+0.005}_{-0.01}$ \\
PG 0953+414   & 0.46  & $0.878^4$     & 0.39           & $0.19 \pm 0.02$        & $0.94 \pm 0.01$\\
Ton 951       & 0.45  & $0.868^4$     & 0.425          & $0.065 \pm 0.04$       & $0.35 \pm 0.05$ \\
\tableline
\end{tabular}
\end{table*}

\citet{shang05} investigated the ultraviolet-to-optical spectral energy distributions of 17 AGNs using quasi-simultaneous spectrophotometry spanning 900-9000\AA\, in rest frame system. They used the observational data from the Far Ultraviolet Spectroscopic Explorer (FUSE), the Hubble Space Telescope (HST) and the 2.1-m telescope at Kitt Peak National Observatory. It is very important that they studied the so-called ''big blue bump'', i.e. the region where the energy output peaks. This detailed investigation allows to cover the wide range in BH mass, Eddington ratio, disk inclination and other parameters, including the range of values for the spectral line luminosity $L_{5100}$. This sample is a heterogeneous one with quite low redshift ($z < 0.5$). All of the optical spectra have been obtained with the 2.1-m telescope at Kitt Peak National Observatory with resolution of $\sim 9$\AA\, in the interval of $\sim 3180-6000$\AA\, and with resolution of $\sim 12$\AA\, in the interval of $\sim 5600-9000$\AA. The special attention has been paid to subtracting the host galaxy contribution. The estimation of the host galaxy contribution is made by \citet{shang05} and appeared to be less than 5\% for all targets.

Near UV spectra of these targets, covering a wavelength range of 1150-3150\AA\, have been obtained from the HST spectroscopic survey. The resulting errors for wavelength of monochromatic radiation are typically less than 1\AA. FUSE far UV spectra are covering observed wavelength of 905-1187\AA\, with the high resolution of $\sim 0.05$\AA. \citet{shang05} also used soft X-ray data collected from the literature \citep{brinkmann97,pfefferkorn01}. These data were obtained from ROST All-Sky Survey. As a result, \citet{shang05} presented detailed energy distribution spectra of these objects and obtained the detailed information on the basic parameters of AGNs, including the values of $FWHM (H_\beta)$, $\lambda L_\lambda(5100 \text{\AA})$, $M_{BH}$ and $l_E = L_{bol} / L_{Edd}$. We used these data to estimate the spin value corresponding to (\ref{eq02}).

The other important factor for the the spin determining is $\mu = \cos{i}$, where $i$ is the inclination angle. The value of $\mu$ can be obtained from the polarimetric data because the degree polarization $P_l(\mu)$ of radiation scattered in the optically thick, but geometrically thin accretion disk depends strongly on the cosine of the inclination angle $\mu$ \citep{chanrdasekhar50,sobolev63}.

The concrete objects from the sample of \citet{shang05} are presented in the Table 1. The detailed data on the important parameters $FWHM(H_\beta)$, $L_{5100}$, $M_{BH}$ and $l_E$ are also presented in the Table 6 of \citet{shang05} paper. Determination of $\mu$ is playing an important part in Eq.(\ref{eq02}). The information on $\mu$ was obtained from polarimetric data from \citet{berriman90,webb93,smith02,afanasiev11,marin16}. The totality of these data allows us to determine the radiation efficiency $\varepsilon (a)$ and then the spin value $a$.

\section{Constraints on the spin value of SMBH}

We demonstrate the determination of constraints on the spin value based on the method, developed by \citet{du14}, for Mrk~290. According to the data, presented by \citet{shang05}, the estimation of the mass of SMBH in Mrk~290 is $M_{BH} = 3.6 \times 10^7 M_{\odot}$, the estimation of the Eddington ratio is $l_E = 0.04$ and monochromatic luminosity $\log{L_{5100}} = 43.31$. Using this data we obtained from Eq.(\ref{eq02}) the following expression for determining the radiation efficiency $\mu^{3/2} = 3.66 \varepsilon (a)$. The value of $\mu$ can be obtained from the polarimetric observations of AGNs presented by \citet{smith02}. According to \citet{smith02} the polarization degree of Mrk~290 continuum is equal to $P_l(\mu) = 0.90\pm 0.04$. This value corresponds to the following value of $\mu$: $\mu = 0.747$ and $\mu^{3/2} = 0.646$. As a result we obtain $\varepsilon (a) = 0.177$ and $a = 0.94$.

Unfortunately, the polarization is not measured for all objects from \citet{shang05} sample. For a number of objects from Table 6 \citep{shang05} the value of $\mu$ was determined through the relation between the virial factor $f$ and estimated mass of the central black hole. This factor is a scale factor that depends on the structure, kinematics and inclination of broad line region (BLR) \citep{collin06}. According to \citet{collin06}, the parameter $f$ is

\begin{equation}
 f = \frac{0.25}{(H/R)^2 + \sin^2{i}},
 \label{eq04}
\end{equation}

\noindent where $H$ is the disk thickness ar the radius $R$. In the standard model of \citet{shakura73} the accretion disk is geometrically thin, i.e. $H/R \ll 0.1$. In this case $f = 0.25 / \sin^2{i}$. For geometrically thin accretion disk the virial relation allows to obtain the following expression for $\sin{i}$:

\begin{equation}
 \sin{i} = \frac{FWHM}{2 c} \left(\frac{R_{BLR}}{R_g}\right)^{1/2},
 \label{eq05}
\end{equation}

\noindent where $FWHM$ is the full width of hydrogen line, $R_g = G M_{BH} / c^2$ is the gravitational radius and $R_{BLR}$ is the radius of the BLR. It is possible to determine the $R_{BLR}$ via the monochromatic luminosity $L_{5100}$. Thus \citet{bentz13} obtained the following expression for $R_{BLR}$:

\begin{equation}
 R_{BLR} = 10^{16.94} \left(\frac{L_{5100}}{10^{44} erg/s}\right)^{0.533}.
 \label{eq06}
\end{equation}

Other useful relation has been obtained by \\ \citet{shen10}:

\begin{equation}
 \sin{i} = 0.198 \left(\frac{FWHM (H_\beta)}{10^3 km/s}\right) \left(\frac{l_E}{M_8}\right)^{1/4}.
 \label{eq07}
\end{equation}

We use the relations (\ref{eq05}) and (\ref{eq06}) for determining the value of $\mu$ for 4C+34.47, IR~07546+3928, Mrk~506, PG~0947+396, PG~1259+593, PG~2349-014, Ton~951.

As a result, the values of the radiation efficiency $\varepsilon (a)$ have been obtained for all objects from the Table 6 of \citet{shang05}. The values of spin for these AGN are presented in the Table 1. For estimates of $\mu$ we used the polarimetric data from \citet{berriman90} (3C~273, 3C~351, PG~0052+254, PG~1100+772, PG~1322+659, PG~1351+640), \citet{smith02} (Mrk~290, Mrk~304, Mrk~509), \citet{afanasiev11} (PG~0953+414, Mrk~509). The inclination angle values of 4C+34.47, IR~07546+3928, Ton~951 and Mrk~506 were estimated via using Eqs. (\ref{eq05})-(\ref{eq07}).

\section{Conclusions}

\citet{shang05} investigated the ultraviolet-to-optical spectral energy distributions of the sample of 17 AGNs using quasi-simultaneous spectropolarimetry spanning 900-9000\AA\, (rest frame) spectroscopic interval. They used the data from the FUSE, HST and 2.1-m telescope at Kitt Peak National Observatory. They compared the behavior of this sample to those of the various thin disk models covering a range in black hole mass, Eddington ratio, disk inclination and other parameters.

We demonstrate that it is possible to estimate the value of a spin of SMBH for AGNs from \citet{shang05} in the framework of the standard geometrically thin accretion disk \citep{shakura73}. The expression (\ref{eq02}), obtained by \citet{du14}, corresponds to this situation. The key moment for estimating SMBH spin value is determining the inclination angle of the accretion disk. The estimate of inclination angle can be obtained from the polarimetric observations in the framework of the theory of radiative transfer of polarized radiation \citep{chanrdasekhar50,sobolev63}. We used for estimate of $\mu$ the results of polarimetric observations obtained by \citet{berriman90,smith02,afanasiev11}. The results of our calculations of the radiative efficiency $\varepsilon (a)$ and determined values of the spin are presented in the Table 1. It should be noted that the majority of SMBHs for AGNs from \citet{shang05} are the strong Kerr SMBHs. The single exclusion is Ton~951 with the spin value $a = 0.35$. This results are very interesting. Our preliminary analysis of the Palomar-Green Catalogue showed the same situation. Majority of SMBHs in AGNs of the PG sample are the Kerr black holes.

\acknowledgments

This research was supported by the Basic Research Program P-7 of Praesidium of Russian Academy of Sciences and the presidential program ''Leading Scientific School-7241.2016.2''. S.D. Buliga was supported by The Committee on Science and Higher School of the Government of St.-Petersburg.

\sloppy
\bibliographystyle{spr-mp-nameyear-cnd}
\bibliography{mybibfile}
\fussy

\end{document}